\documentclass[prl,aps,twocolumn, floatfix, superscriptaddress,showpacs]{revtex4}
\usepackage{amsmath,bm,graphicx}
\usepackage{amssymb}
\usepackage{amsfonts}

\newcommand{\V}[1]{{\bf #1}}

\newcommand{\vv}[1]{{\mathbf #1}}

\begin{document}
\title{The life-cycle of drift-wave turbulence driven by small scale instability}
\author{Colm Connaughton}
\affiliation{Mathematics Institute, University of Warwick, Coventry CV4 7AL, UK}
\affiliation{Centre for Complexity Science, University of Warwick, Coventry CV4 7AL, UK}
\author{Sergey Nazarenko}
\affiliation{Mathematics Institute, University of Warwick, Coventry CV4 7AL, UK}
\author{Brenda Quinn}
\affiliation{Mathematics Institute, University of Warwick, Coventry CV4 7AL, UK}

\date{\today}

\begin{abstract}
We demonstrate theoretically and numerically the  zonal-flow/drift-wave
feedback mechanism for the LH transition in an idealised model of plasma turbulence
driven by a small scale instability. Zonal flows are generated
by  a secondary modulational instability of the modes which are directly driven by the
primary instability. The zonal flows then suppress the small scales thereby arresting
the energy injection into the system,
a process which can be described  using  nonlocal wave turbulence
theory. Finally, the arrest of the energy input results in saturation of the
zonal flows  at a level which can be estimated
from the theory and the system reaches stationarity without damping of the large scales.
\end{abstract}
\maketitle

Designing a device capable of controlled, self-sustained
nuclear fusion is one of the principal technological challenges of the
21st century.  Attention is now heavily
focused on the construction of the International Thermonuclear Test Reactor (ITER)
which aims to demonstrate the feasibility of controlled fusion
for the first time. Of the myriad of engineering challenges posed by ITER,
one of the most critical, the problem of magnetic plasma confinement, relates
intimately to foundational issues in turbulence theory, the topic of this Letter.
One difficulty of magnetic plasma confinement is due to instabilities
(such as the ion temperature gradient, drift-dissipative
and  electron temperature gradient instabilities). These generate waves which
interact to produce wave turbulence thereby greatly enhancing transport.
Heat and plasma then diffuse more quickly from the hot, dense core to the
periphery resulting in a loss of confinement.  Common features of these
instabilities are that they are small scale, typically of the order of the ion
gyro-radius or below, and principally generate meridional waves
(with wave crests along the minor radius of the torus).

Significant progress in the confinement problem came with the experimental
discovery \cite{WAG1982abbrev} that, under the right conditions, the turbulent transport coefficient
suddenly drops.
This effect, called the Low-to-High (LH) confinement transition,
can be exploited to design more efficient confinement.  Although traces of the LH transition were
replicated in numerical simulations \cite{NBD2007}, including full gyro-kinetic calculations \cite{VIL2004abbrev},
the underlying mechanism was initially poorly understood since realistic numerical
simulations of tokamaks are universally incapable of resolving the full range of
scales involved. It is now generally accepted \cite{DIIH2005} that the LH transition is triggered
by the self-organisation of coherent zonal flows (ZF's) in the poloidal direction which
act as transport barriers since they act to shear away turbulent fluctuations which
try to cross them. Current research is heavily focused on understanding
and parameterising these ZF's and their
interaction with the drift wave turbulence at the experimental, numerical and
theoretical levels.

In this Letter we present a simple theoretical model of the confinement
mechanism described above. Our model uses the Charney-Hasegawa-Mima (CHM) equation
\cite{HM1977}, the simplest
single-fluid description of a strongly magnetised plasma. This equation
is rarely used in modern plasma research because drastic simplifications are
required to obtain it from the true dynamical equations. On the other hand,
simple models often repay in terms of analytical tractability and conceptual
clarity what they lose in terms of direct applicability.
This is certainly the
case here. Our model gives a semi-analytical description of the full life-cycle
of turbulence generated by a small-scale primary instability. Before presenting details,
we outline the main results. Meridional waves - drift waves in the
case of the CHM model - are generated by a primary instability at the scale of the ion
gyro-radius. The subsequent evolution has three main stages. In stage 1,
waves  grow  until they generate seeds of a ZF via a secondary
modulational instability \cite{MN1994,OPSSS2004,CNNQ2010}.  The ZF is large-scale
compared to the waves. In stage 2, the ZF's grows by {\em direct}
interaction with the small scales rather than by a Kraichnan-style inverse cascade
which would involve local transfer of energy between scales.  This is because
the nonlocal interaction in drift wave turbulence is intrinsic, i.e.
it typically develops in time even for broadband initial spectra.
However, the fact that the primary instability produces a relatively narrow
 initial drift-wave spectrum helps nonlocality.
Due to nonlocality, the kinetic energy of the
small scales is not conserved as it would be in a local cascade. In fact, the
growing ZF {\em removes} energy from the turbulence since the combined
energy of large and small scales is conserved.  The wave turbulence which
gave birth to the ZF in stage 1 is thus suppressed by its growth.  In stage
3, this negative feedback reaches a steady state in which suppression of
turbulence by the large-scale
ZF exactly balances its generation by the small
scale instability. We note that Eq.~(\ref{eq:CHM}) also has a  geophysical interpretation
which we do not discuss (see references in \cite{CNNQ2010}).

The CHM model does not contain small scale instabilities of the kind mentioned above. In what
follows, we {\em model} these  instabilities with a multiplicative linear forcing
term applied to selected wavenumbers. We solve the following equation on a plane with the $x$-axis along
the plasma density gradient $\nabla n_0$ and the $y$-axis transverse to this:
\begin{equation}
\label{eq:CHM}
\left( \partial_ t -{\mathcal L} \right)
\left(\rho^2 \Delta \psi - \psi  \right) + v_* \, \partial_y \psi - \rho^2 J[\psi, \Delta \psi]= 0.
\end{equation}
Here $\psi(x,y,t)$ is the streamfunction, $\Delta$ is the 2D Laplacian,
$J[\psi, \Delta \psi] = \partial_x \psi \partial_y \Delta \psi - \partial_y \psi \partial_x \Delta \psi$
and ${\mathcal L}$ is a linear operator having Fourier-space representation
${\mathcal L}_{\V{k}} = \gamma_\V{k} -\nu_m k^{2 m}$ with $\gamma_\V{k}>0$
modeling the instability and $\nu_m$ a hyperviscosity coefficient.
Here, $\rho$ is the ion Larmor radius at
the electron temperature, $v_* =\omega_{B} \rho^2/L_n $ is the drift velocity,
$\omega_{B}$ is the ion gyro-frequency and $L_n = n_0/|\nabla n_0|$ is the density profile length-scale. 
$\rho$ and $v_*$ are both taken to be constant.
For the instability forcing, $\gamma_\V{k}$, we seek the simplest form which
retains the key features of the relevant plasma instabilities mentioned above,
namely that they act at small scales and primarily generate meridional waves.
For these purposes it suffices to take
$\gamma_{\V{k}} = \gamma\, \delta_{{\bf k},{\bf k}_i}$, with
$\V{k}_i=(0,\rho^{-1})$ and $\gamma>0$, which forces a single meriodinal mode at the scale of the gyro-radius.
In principle, a more realistic form for $\gamma_\V{k}$ can be obtained
by considering the linear dynamics of a higher level model which {\em does}
contain an intrinsic instability such as the Hasegawa-Wakatani (HW) model in
the case of the drift-dissipative instability \cite{NBD2007}. The dimensionless parameter,
$\chi = \frac{\gamma \rho}{ v_*}$,
is a measure of nonlinearity of the system assuming
that the linear instability saturates at a level determined by the amplitude of the
nonlinear term in Eq.(\ref{eq:CHM}).
Eq.~(\ref{eq:CHM}) has exact plane wave solutions
$\psi = \exp(-i (\omega_{\bf k} t - {\bf k} \cdot {\bf x}) + \gamma_{\bf k} t -\nu_m k^{2m} t )$ with  dispersion relation,
$\omega_{\bf k} = -v_* k_y/ (1 + \rho^2 k^2)$.

\begin{figure}
\includegraphics[width=6.5cm] {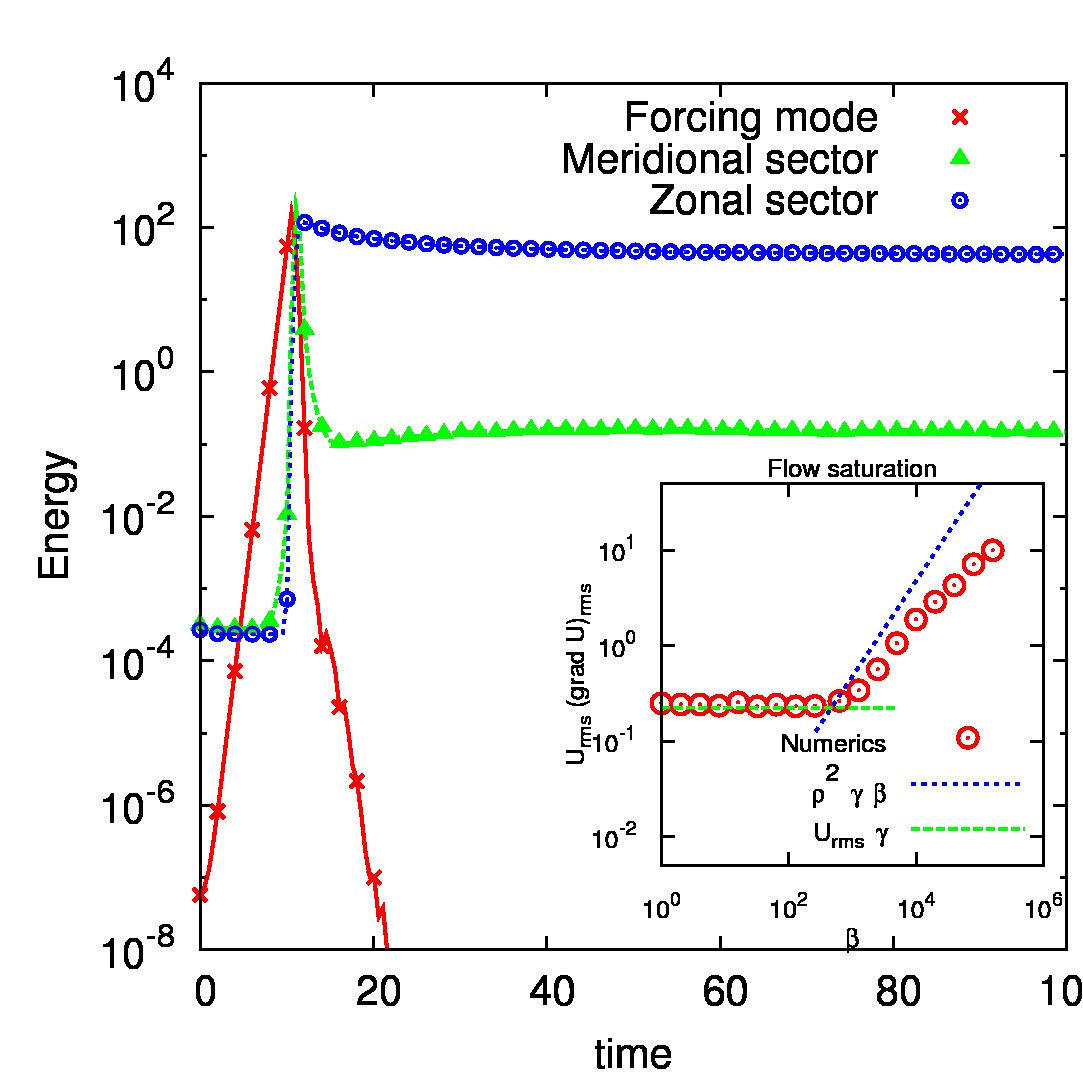}
\caption{\label{fig-Chi3.75E-4}
(color online) Energy contained in the forced mode (crosses),
the zonal sector (circles) and meridional sector
(triangles) as functions of time for  $\chi=3.75\times10^{-4}$. The inset shows the
saturation level of the ZF as a function of $\beta$.}
\end{figure}

\begin{figure}[th]
\begin{center}
\includegraphics[width=7cm]{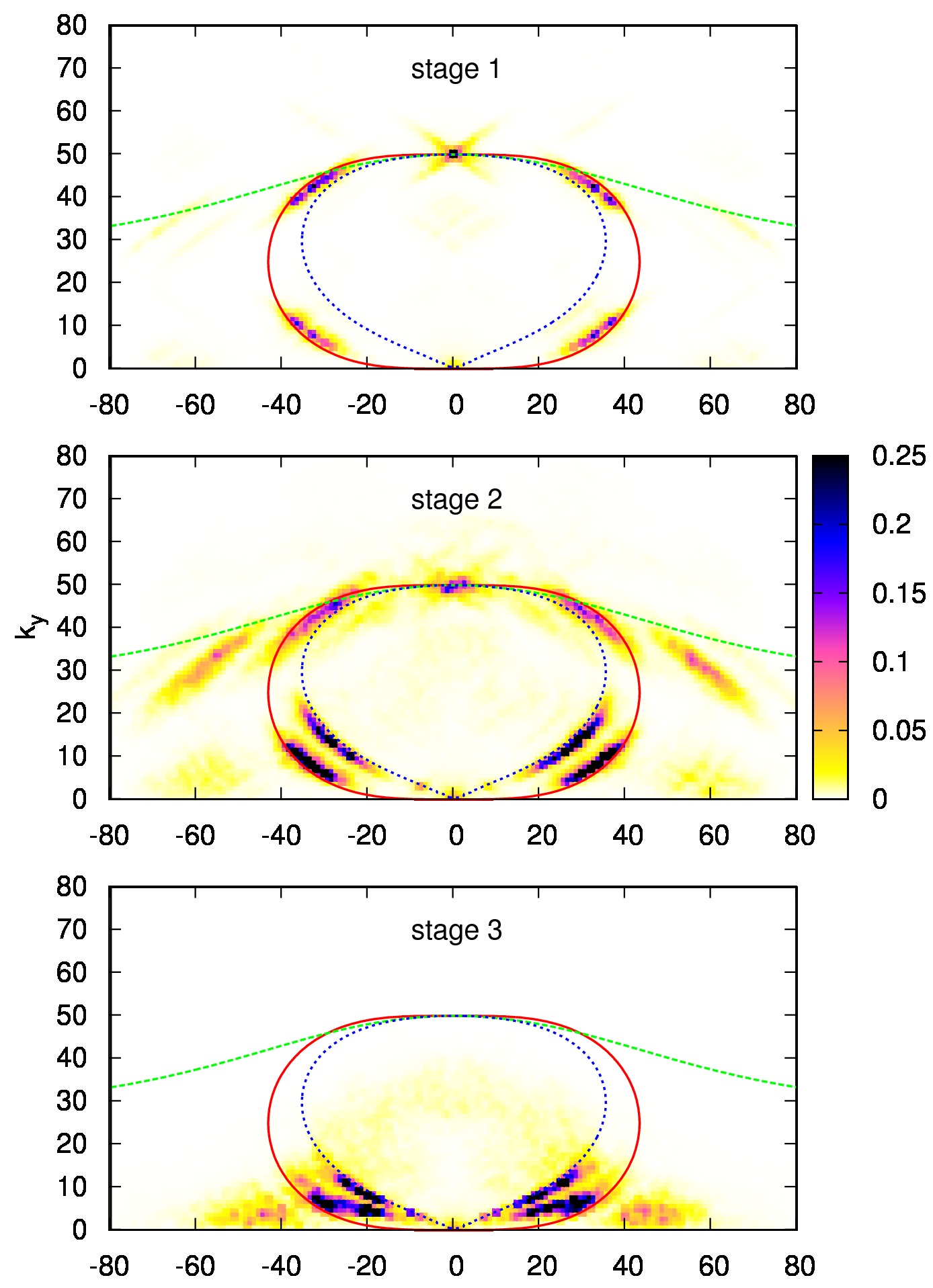}
\end{center}
\caption{\label{fig:NL2e-3_Ek}(color online) Snapshots of the 2D energy spectrum at times  $11.0$, $11.5$ and $20$ for the case $\chi=3.74\times 10^{-4}$}
\end{figure}

Without instability or dissipation (${\mathcal L} \equiv 0$),
 Eq.~(\ref{eq:CHM}) has two invariants:
$E = \frac 1{2} \int [\rho^{-2}\psi^2 + (\nabla \psi )^2] \, d {\bf x}$ and
$\Omega  = \frac 1{2} \int [ \rho^{-2}(\nabla \psi )^2 + (\nabla^2 \psi )^2] \, d {\bf x}$.
These are closly related to the energy and enstrophy in the 2D 
Euler equations. 2D Euler turbulence exhibits an inverse cascade
of energy from small scales to large. Likewise, CHM turbulence transfers energy 
to large scales. In the weakly nonlinear limit, however, CHM turbulence differs
from 2D Euler turbulence in some important respects. Firstly, in this
limit, the energy transfer turns out to be nonlocal in scale \cite{BNZ1990} meaning
that it is dominated by interactions between the smallest and largest scales. 
Secondly, there is an additional statistical invariant, the {\em zonostrophy}
\cite{BNZ1991}, which renders the transfer anisotropic \cite{BNZ1991,NQ2009}.  
The theory of nonlocal drift wave turbulence was first worked out in 
\cite{BNZ1990,NAZ1991} and a detailed summary can be found in
\cite{CNQ2010B}.  In this Letter, we will present a self-consistent but 
qualitative physical argument to arrive at the same results.

Consider a drift wavepacket with wavenumber $\vv{k}$ moving on a background 
consisting of a large-scale ZF with typical wavenumber $\vv{p}$. The ZF need not 
be perfect and can have finite $p_x$, but we insist that it is large scale: 
$p \ll k$ and $p \ll 1/\rho$ so that a the  WKB approximation is 
appropriate \cite{DNZ1992}.  The ZF is stationary in a frame moving in 
the $y$-direction with the drift velocity $v_*$ \cite{CNQ2010B}.  As a consequence 
of the WKB approximation, the wave packet frequency must be conserved in this 
frame:
\begin{equation}
\label{eq:om1}
 v_* k_y  + \omega_\vv{k} =
\frac{v_* k_y \rho^2 k^2}{1+\rho^2 k^2} = k^2\,\omega_\vv{k} = v= \mbox{const}.
\end{equation}
Hence, as $\vv{k}$ changes due to distortion of the wavepacket by the 
ZF,  it is restricted to remain on the 1D curve in the 2D $\vv{k}$-space given 
by \eqref{eq:om1}, hereafter called a $v$-curve. A typical $v$-curve
is indicated by the green (dashed) line in Fig.~\ref{fig:NL2e-3_Ek}.  According to 
WKB theory, it is waveaction of the small scales, $n_\vv{k} = e_\vv{k}/\omega_\vv{k}$,  which is 
conserved rather than the energy, $e_\vv{k}$. On each $v$-curve, 
$n_\vv{k} \sim k^2 e_\vv{k}$ from Eq.~(\ref{eq:om1}).  Since $k^2$ increases as a wavepacket 
moves along a $v$-curve from the source (at the $k_y$-axis) to the sink (at high 
$k$'s),  $e_\vv{k}$  must decrease to keep $n_\vv{k}$ constant. The energy thus lost 
is transferred to the ZF.

There are two qualitatively different regimes corresponding to weak and strong 
ZFs. The weak regime occurs when drift wave packets
live long enough to cross many ZF oscillations, experiencing multiple random 
distortions and executing a random walk in $\bf k$-space along the $v$-curves. 
It can be described by a diffusion equation:
\begin{equation}
\label{eq:spectrum_evolution}
\frac{\partial n_\vv{k}}{\partial t} =  \partial_v \left( D_\vv{k} \partial_v n_\vv{k} \right) +\gamma_\vv{k} n_\vv{k}
-\nu_m k^{2m}  n_\vv{k},
\end{equation}
where $\partial_v$ is the derivative {\em along} the $v$-curve. A detailed quantitative derivation of 
Eq.~(\ref{eq:spectrum_evolution}) is presented in \cite{CNQ2010B}. 
The diffusion coefficient, $D_\vv{k}$, is  proportional to
the intensity of the  large-scale ZF, and can be estimated as
$D_\vv{k} \sim (\dot k_x)^2 \tau,
$
where $\dot k_x \sim - k_y S$ is the rate of change of $k_x$ due to the ZF shear
$S=\partial_x U \sim p U$,
 and  $\tau$ is the correlation time
$\tau= {1 /( p |{\dot x}|)} =  {1 /(p |{\partial_{k_x} \omega_k}|)} \sim
{1  /( p  v_*) }.
$
In these estimates we took into account that the essential dynamics happens at the
scales $k_x \sim k_y \sim 1/\rho$.
Thus
\begin{equation}
D_\vv{k} \sim
\frac{ S^2}  {\rho^2 v_* p  }  \sim \frac{ SU }{\rho^2 v_*   }.
\label{d2}
\end{equation}

For fixed $D_\vv{k}$, Eq.~(\ref{eq:spectrum_evolution}) indicates that $n_\vv{k}$ 
evolves independently on each $v$-curve and is controlled by the 
maximum value of $\lambda_\vv{k} = \gamma_\vv{k}-\nu_m k^{2m}$ on the curve. If 
$\lambda_\vv{k}$ is positive, $n_\vv{k}$ will initially grow exponentially.  
Correspondingly, if $\lambda_\vv{k}$ is negative,  $n_\vv{k}$ will be 
exponentially damped.  Large positive $\lambda_\vv{k}$ are obtained on curves 
passing close to the maximum of $\gamma_k$. In our case this is the forcing mode, 
$\vv{k} =(0,k_i)$.  Negative values are found away from this maximum, where 
dissipation dominates.
The growth of $n_\vv{k}$ on a $v$-curve with  positive $\lambda_\vv{k}$  cannot,
however, continue indefinitely. Diffusion along a curve transfers energy to large
scales. This in turn increases $D_\vv{k}$  so that the growth due to positive
$\lambda_\vv{k}$ can begin to be tempered by increasingly rapid diffusion of 
wavepackets from the instability region to the dissipation region. The region
of growing $n_\vv{k}$ shrinks until a steady state is reached (stage 3).
Those $v$-curves which originally had  positive growth reduce their
growth rates to zero and the spectrum on them dies out.
The large-scale ZF in turn saturates at a finite intensity determined by
balancing the forcing and the diffusion terms on the right hand side of
Eq.~(\ref{eq:spectrum_evolution}),  giving
$ D_\vv{k} \rho^2 \sim \gamma_\mathrm{max}.
$
 Eq. \eqref{d2} gives the saturation criterion:
 \begin{equation}
 \label{Usatur}
 U_\mathrm{rms} S_\mathrm{rms} \sim  v_*
 \gamma_\mathrm{max}.
 \end{equation}
where rms denotes root-mean square.
Counterintuitively, the saturation velocity increases with $v_*$ as the 
flow becomes more linear.
Eq.~(\ref{Usatur}) fails for large $\chi$ when turbulence is strong and the 
$v_*$-term in Eq.~(\ref{eq:CHM})can be neglected.
In the strong turbulence regime, a wavepacket cannot cross even a single ZF 
oscillation since rapid distortion by the zonal shear brings its 
wavenumber to the dissipation scale.  This gives the saturation criterion
\begin{equation}
 \label{Usatur-bis}
  S_\mathrm{rms} \sim   \gamma_\mathrm{max}.
 \end{equation}
Comparing Eqs.~(\ref{Usatur}) and \eqref{Usatur-bis},  the crossover 
between the two regimes occurs when $ \gamma_\mathrm{max} \sim v_* p$ 
or when $U_\mathrm{rms} \sim v_*$.  The estimate \eqref{Usatur-bis}, previously put 
forward in \cite{KW1994}, is valid only for strong ZF's with $U_\mathrm{rms}  \gtrsim v_*$.

We performed numerical simulations using a standard
pseudo-spectral code at a resolution of $256\times 256$ with $m=8$,
$\nu_m=1.0\times 10^{-29}$, $\rho=0.02$ and $\gamma=1.2$.
The initial condition was 
spatial white noise of very low amplitude.  Fig.~\ref{fig-Chi3.75E-4} summarizes
results obtained  with $\beta=1.6\times 10^{5}$ giving
$\chi=3.75\times10^{-4}$.  We divided $\V{k}$-space into a zonal sector,
$\left|k_x\right|>\left|k_y\right|$, and a meridional sector, 
$\left|k_x\right|\leq \left|k_y\right|$ (excluding the forced mode).
Fig.~\ref{fig-Chi3.75E-4}
shows the energy of each sector and of the forced mode as functions
of time.
Initially, the forced mode
grows exponentially, and the zonal and meridional energies remain
at a level set by the initial noise. This is followed
by a sharp growth of the the zonal and the off-zonal energies once the amplitude
of the forced mode gets high enough to facilitate nonlinear mode coupling (stage
1).
 Next, the meridional modes, including the
forced mode itself,  are suppressed while the zonal modes retain their energy (stage 2).
 Finally a steady state is reached (stage 3) in which the instability
forcing is  switched off and the majority of the energy is zonal.

Fig.~\ref{fig:NL2e-3_Ek} shows the time evolution of the 2D energy spectrum in
$\bf k$-space normalized by the maximum.
The forced mode initially dominates but
such waves are modulationally unstable \cite{MN1994,OPSSS2004,CNNQ2010}.
The first indication
of the nonlinear evolution seen in stage 1 is therefore the growth from the background noise of
those modes whose modulational instability growth rate is greatest.
In the limit of weak nonlinearity, the modulationally unstable modes concentrate
on the resonant curve passing through the forcing mode,
$\omega(\vv{k}) = \omega(\vv{k}_1) +\omega(\vv{k}_2)$ with
$\vv{k} = \vv{k}_1 +\vv{k}_2$ for $\vv{k_3}=(k_i,0)$ \cite{CNNQ2010}.
The maximally unstable perturbations of a meridional wave have a significant zonal
component (see Fig. 2 of \cite{CNNQ2010}). These are visible, along with
the resonant curve,  in the left panel of Fig.~\ref{fig:NL2e-3_Ek}.
They are the seeds of the ZF's responsible for the
LH transition.
The very rapid decrease in meridional turbulent energy observed in
Fig.~\ref{fig-Chi3.75E-4} at stage 2 is related with the ZF-induced diffusion
along the $v$-curves with a diffusion coefficient growing as the ZF is becoming stronger.
 This diffusion process is clearly visible in the
middle frame of Fig.~\ref{fig:NL2e-3_Ek}. Deviations from the theoretical $v$-curve
can be attributed to absence of a large scale separation between the forcing and
the zonal modes assumed in the derivation of Eq.~(\ref{eq:spectrum_evolution}).
We remark that if there is a nonlocal interaction with {\em small-scale} ZF's
then diffusion proceeds along different curves, the level sets
of zonostrophy \cite{NAZ1991,CNQ2010B}. One such curve, passing through the forced mode is
shown in Fig.~\ref{fig:NL2e-3_Ek} by the blue dashed line. A secondary ZF
branch emerges on this curve, a point which we hope to explore in a future publication.

At stage 3,
the zonal flow saturation estimates Eqs.~(\ref{Usatur}) and (\ref{Usatur-bis}) are in reasonable agreement with
the numerical measurements in the inset of Fig.~\ref{fig-Chi3.75E-4}. The
saturation values of $ U_\mathrm{rms} S_\mathrm{rms}$ are compared against the estimated values
over a range of $v_*$ values. For low $v_*$  there is good agreement with the
strong turbulence estimate, Eq.~(\ref{Usatur-bis}). For large $v_*$ there is
reasonable agreement with the weak turbulence estimate  Eq.~(\ref{Usatur})
(the discrepancy is a factor of 2 -- 3).  The departure at large $v_*$
from the prediction $ U_\mathrm{rms} S_\mathrm{rms} \propto v_*$ is likely due to the fact
that the scale separation becomes less pronounced for large  $v_*$.

To conclude, we have analyzed the zonal-flow/drift-wave feedback loop mechanism
for the LH transition in an idealised model of plasma turbulence driven by
a small scale instability. The process has 3 stages:
(i) ZF's are generated by a modulational instability of the small-scale modes
which are directly excited by the instability, (ii) the ZF's
feed back on the small-scale modes and suppress them via a nonlocal
mechanism which closely follows the scenario predicted in
\cite{BNZ1990,NAZ1991} , (iii) the energy injection into the system stops
and the zonal energy saturates at a level which is in reasonable agreement with the
theoretical estimates, Eqs.~(\ref{Usatur}) and (\ref{Usatur-bis}),
obtained from the nonlocal weak and strong turbulence
theories
respectively.

The ZF saturation mechanism described in this Letter is remarkable because it
is achieved entirely dynamically without the need to remove the small scale
driving instability or to invoke a dissipation mechanism for
the large scales. The fact that it works transparently in a simple model like 
the CHM model is encouraging for more realistic applications since 
it suggests that the mechanism is quite robust. We conclude with some physical 
estimates  aiming
to establish the extent to which this mechanism could, in principle, be observable 
in real tokamaks.
The strong turbulence scaling Eq.~(\ref{Usatur-bis}), appropriate for $U \gtrsim v_*$,
has been widely used in tokamak theory since \cite{KW1994}. On the other hand, 
the weak turbulence estimate (\ref{Usatur}) is new. It should be used when 
$U \lesssim v_*$.  Zonal flows in the DIIID  and JET tokamaks have been reported 
with $U \sim 5-35$ km/sec \cite{SOL2004, CRO2006} and $v_*$ for both systems ranges from $5$ km/sec in the core to  
$100$ km/sec in the pedestal region.  Both the weak and the strong regimes 
should therefore be realisable.  Some idea of the feasibility of 
the weak vs strong saturation regimes can be obtained for the most typical 
instability in the tokamak plasma core, the 
Ion-Temperature Gradient instability for which $\gamma_\mathrm{max} = v_T/\sqrt{R L_T}$, 
where $R$ is the major radius and $L_T$ is the temperature gradient length.
The weak/strong crossover occurs when $\gamma_\mathrm{max}/p \sim v_*$, i.e.
when $ \rho p \sim  L_n \sqrt{R L_T}$ where $L_n$ is the typical lenght scale of
the density gradient.
For ZFs which appear inside pedestal regions, $L_n \sim L_T $ and 
$\rho p \sim 0.1$. The crossover condition is thus $R/L_n \sim 100$.
In large tokamaks $R \gtrsim 1$ m and in the pedestal region $L_n \gtrsim 1$ cm.
Thus, both the weak and the strong regimes of the ZF generation can be expected.
For the edge region, one should use another estimate for $\gamma_{max}$ which is
based on the drift-dissipative instability. In this case one can show that the weak
regime should be expected when the electron mean free path exceeds $R$.


\begin{thebibliography}{17}
\expandafter\ifx\csname natexlab\endcsname\relax\def\natexlab#1{#1}\fi
\expandafter\ifx\csname bibnamefont\endcsname\relax
  \def\bibnamefont#1{#1}\fi
\expandafter\ifx\csname bibfnamefont\endcsname\relax
  \def\bibfnamefont#1{#1}\fi
\expandafter\ifx\csname citenamefont\endcsname\relax
  \def\citenamefont#1{#1}\fi
\expandafter\ifx\csname url\endcsname\relax
  \def\url#1{\texttt{#1}}\fi
\expandafter\ifx\csname urlprefix\endcsname\relax\def\urlprefix{URL }\fi
\providecommand{\bibinfo}[2]{#2}
\providecommand{\eprint}[2][]{\url{#2}}

\bibitem[{\citenamefont{Wagner et~al.}(1982)}]{WAG1982abbrev}
\bibinfo{author}{\bibfnamefont{F.}~\bibnamefont{Wagner}} \bibnamefont{et~al.},
  \bibinfo{journal}{Phys. Rev. Lett.} \textbf{\bibinfo{volume}{49}},
  \bibinfo{pages}{1408} (\bibinfo{year}{1982}).

\bibitem[{\citenamefont{{Numata} et~al.}(2007)\citenamefont{{Numata}, {Ball},
  and {Dewar}}}]{NBD2007}
\bibinfo{author}{\bibfnamefont{R.}~\bibnamefont{{Numata}}},
  \bibinfo{author}{\bibfnamefont{R.}~\bibnamefont{{Ball}}}, \bibnamefont{and}
  \bibinfo{author}{\bibfnamefont{R.~L.} \bibnamefont{{Dewar}}},
  \bibinfo{journal}{Phys. Plasmas} \textbf{\bibinfo{volume}{14}},
  \bibinfo{pages}{102312} (\bibinfo{year}{2007}).

\bibitem[{\citenamefont{Villard et~al.}(2004)}]{VIL2004abbrev}
\bibinfo{author}{\bibfnamefont{L.}~\bibnamefont{Villard}} \bibnamefont{et~al.},
  \bibinfo{journal}{Nucl. Fusion} \textbf{\bibinfo{volume}{44}},
  \bibinfo{pages}{172} (\bibinfo{year}{2004}).

\bibitem[{\citenamefont{{Diamond} et~al.}(2005)\citenamefont{{Diamond}, {Itoh},
  {Itoh}, and {Hahm}}}]{DIIH2005}
\bibinfo{author}{\bibfnamefont{P.~H.} \bibnamefont{{Diamond}}},
  \bibinfo{author}{\bibfnamefont{S.-I.} \bibnamefont{{Itoh}}},
  \bibinfo{author}{\bibfnamefont{K.}~\bibnamefont{{Itoh}}}, \bibnamefont{and}
  \bibinfo{author}{\bibfnamefont{T.~S.} \bibnamefont{{Hahm}}},
  \bibinfo{journal}{Plasma Phys. Control. Fusion}
  \textbf{\bibinfo{volume}{47}}, \bibinfo{pages}{R35} (\bibinfo{year}{2005}).

\bibitem[{\citenamefont{Hasegawa and Mima}(1977)}]{HM1977}
\bibinfo{author}{\bibfnamefont{A.}~\bibnamefont{Hasegawa}} \bibnamefont{and}
  \bibinfo{author}{\bibfnamefont{K.}~\bibnamefont{Mima}},
  \bibinfo{journal}{Phys. Rev. Lett.} \textbf{\bibinfo{volume}{39}},
  \bibinfo{pages}{205} (\bibinfo{year}{1977}).

\bibitem[{\citenamefont{{Manin} and {Nazarenko}}(1994)}]{MN1994}
\bibinfo{author}{\bibfnamefont{D.~Y.} \bibnamefont{{Manin}}} \bibnamefont{and}
  \bibinfo{author}{\bibfnamefont{S.~V.} \bibnamefont{{Nazarenko}}},
  \bibinfo{journal}{Phys. Fluids} \textbf{\bibinfo{volume}{6}},
  \bibinfo{pages}{1158} (\bibinfo{year}{1994}).

\bibitem[{\citenamefont{Onishchenko et~al.}(2004)\citenamefont{Onishchenko,
  Pokhotelov, Sagdeev, Shukla, and Stenflo}}]{OPSSS2004}
\bibinfo{author}{\bibfnamefont{O.}~\bibnamefont{Onishchenko}},
  \bibinfo{author}{\bibfnamefont{O.}~\bibnamefont{Pokhotelov}},
  \bibinfo{author}{\bibfnamefont{R.}~\bibnamefont{Sagdeev}},
  \bibinfo{author}{\bibfnamefont{P.}~\bibnamefont{Shukla}}, \bibnamefont{and}
  \bibinfo{author}{\bibfnamefont{L.}~\bibnamefont{Stenflo}},
  \bibinfo{journal}{Nonlin. Proc. Geophys.} \textbf{\bibinfo{volume}{11}},
  \bibinfo{pages}{241} (\bibinfo{year}{2004}).

\bibitem[{\citenamefont{{Connaughton}
  et~al.}(2010{\natexlab{a}})\citenamefont{{Connaughton}, {Nadiga},
  {Nazarenko}, and {Quinn}}}]{CNNQ2010}
\bibinfo{author}{\bibfnamefont{C.}~\bibnamefont{{Connaughton}}},
  \bibinfo{author}{\bibfnamefont{B.}~\bibnamefont{{Nadiga}}},
  \bibinfo{author}{\bibfnamefont{S.~V.} \bibnamefont{{Nazarenko}}},
  \bibnamefont{and} \bibinfo{author}{\bibfnamefont{B.~E.}
  \bibnamefont{{Quinn}}}, \bibinfo{journal}{J. Fluid Mech.}
  \textbf{\bibinfo{volume}{654}}, \bibinfo{pages}{207}
  (\bibinfo{year}{2010}{\natexlab{a}}).

\bibitem[{\citenamefont{{Balk} et~al.}(1990)\citenamefont{{Balk}, {Nazarenko},
  and {Zakharov}}}]{BNZ1990}
\bibinfo{author}{\bibfnamefont{A.~M.} \bibnamefont{{Balk}}},
  \bibinfo{author}{\bibfnamefont{S.~V.} \bibnamefont{{Nazarenko}}},
  \bibnamefont{and} \bibinfo{author}{\bibfnamefont{V.~E.}
  \bibnamefont{{Zakharov}}}, \bibinfo{journal}{Phys. Lett. A}
  \textbf{\bibinfo{volume}{146}}, \bibinfo{pages}{217} (\bibinfo{year}{1990}).

\bibitem[{\citenamefont{{Balk} et~al.}(1991)\citenamefont{{Balk}, {Nazarenko},
  and {Zakharov}}}]{BNZ1991}
\bibinfo{author}{\bibfnamefont{A.~M.} \bibnamefont{{Balk}}},
  \bibinfo{author}{\bibfnamefont{S.~V.} \bibnamefont{{Nazarenko}}},
  \bibnamefont{and} \bibinfo{author}{\bibfnamefont{V.~E.}
  \bibnamefont{{Zakharov}}}, \bibinfo{journal}{Phys. Lett. A}
  \textbf{\bibinfo{volume}{152}}, \bibinfo{pages}{276} (\bibinfo{year}{1991}).

\bibitem[{\citenamefont{Nazarenko and Quinn}(2009)}]{NQ2009}
\bibinfo{author}{\bibfnamefont{S.}~\bibnamefont{Nazarenko}} \bibnamefont{and}
  \bibinfo{author}{\bibfnamefont{B.}~\bibnamefont{Quinn}},
  \bibinfo{journal}{Phys. Rev. Lett.} \textbf{\bibinfo{volume}{103}},
  \bibinfo{pages}{118501} (\bibinfo{year}{2009}).

\bibitem[{\citenamefont{{Nazarenko, S.V.}}(1991)}]{NAZ1991}
\bibinfo{author}{\bibnamefont{{Nazarenko, S.V.}}}, \bibinfo{journal}{JETP
  Lett.} \textbf{\bibinfo{volume}{53}}, \bibinfo{pages}{628}
  (\bibinfo{year}{1991}).

\bibitem[{\citenamefont{{Connaughton}
  et~al.}(2010{\natexlab{b}})\citenamefont{{Connaughton}, {Nazarenko}, and
  {Quinn}}}]{CNQ2010B}
\bibinfo{author}{\bibfnamefont{C.}~\bibnamefont{{Connaughton}}},
  \bibinfo{author}{\bibfnamefont{S.~V.} \bibnamefont{{Nazarenko}}},
  \bibnamefont{and} \bibinfo{author}{\bibfnamefont{B.~E.}
  \bibnamefont{{Quinn}}} (\bibinfo{year}{2010}{\natexlab{b}}),
  \bibinfo{note}{arXiv:1012.2714v1 [nlin.CD]}.

\bibitem[{\citenamefont{{Dyachenko} et~al.}(1992)\citenamefont{{Dyachenko},
  {Nazarenko}, and {Zakharov}}}]{DNZ1992}
\bibinfo{author}{\bibfnamefont{A.}~\bibnamefont{{Dyachenko}}},
  \bibinfo{author}{\bibfnamefont{S.~V.} \bibnamefont{{Nazarenko}}},
  \bibnamefont{and} \bibinfo{author}{\bibfnamefont{V.~E.}
  \bibnamefont{{Zakharov}}}, \bibinfo{journal}{Phys. Lett. A}
  \textbf{\bibinfo{volume}{165}}, \bibinfo{pages}{330} (\bibinfo{year}{1992}).

\bibitem[{\citenamefont{{Kingsbury} and {Waltz}}(1994)}]{KW1994}
\bibinfo{author}{\bibfnamefont{O.~T.} \bibnamefont{{Kingsbury}}}
  \bibnamefont{and} \bibinfo{author}{\bibfnamefont{R.~E.}
  \bibnamefont{{Waltz}}}, \bibinfo{journal}{Phys. Plasmas}
  \textbf{\bibinfo{volume}{1}}, \bibinfo{pages}{2319} (\bibinfo{year}{1994}).

\bibitem[{\citenamefont{Solomon et~al.}(2004)}]{SOL2004}
\bibinfo{author}{\bibfnamefont{W.~M.} \bibnamefont{Solomon}}
  \bibnamefont{et~al.}, \bibinfo{journal}{Rev. Sci. Instrum.}
  \textbf{\bibinfo{volume}{75}} (\bibinfo{year}{2004}).

\bibitem[{\citenamefont{{Cromb\'e}}(2006)}]{CRO2006}
\bibinfo{author}{\bibfnamefont{K.}~\bibnamefont{{Cromb\'e}}}, in
  \emph{\bibinfo{booktitle}{33rd EPS Conference on Plasma Phys.}}
  (\bibinfo{year}{2006}), pp. \bibinfo{pages}{30I D--5.015}.

\end{thebibliography}

\end{document}